\documentclass[11pt]{article}
\usepackage{graphicx}

\newcommand{\BABARPubYear}    {06}

\newcommand{\BABARConfNumber} {021}

\input babarsym

\long\def\inst#1{\par\nobreak\kern 4pt\nobreak
    {\it #1}\par\vskip 10pt plus 3pt minus 3pt}

\begin{document}
{\pagestyle{empty}

\begin{flushright}
\babar-CONF-\BABARPubYear/\BABARConfNumber \\

July 2006 \\
\end{flushright}

\par\vskip 5cm

\begin{center}
\Large \bf 
Measurement of the Mass and Width and Study of the Spin of the $\Xi(1690)^0$ Resonance from $\Lambda_c^+ \rightarrow \Lambda \bar K^0 K^+$ Decay at \babar
\end{center}
\bigskip

\begin{center}
\large The \babar\ Collaboration\\
\mbox{ }\\
\today
\end{center}
\bigskip \bigskip

\begin{center}
\large \bf Abstract
\end{center}
The $\Xi(1690)^{0}$ resonance is
observed in the $\Lambda \bar K^0$ channel in the decay
$\Lambda_c^+ \rightarrow \Lambda \bar K^0 K^+$,
from a data sample corresponding to a total integrated luminosity of $\sim$ 200 fb$^{-1}$ 
recorded by the \babar\ detector at the \pep2\ asymmetric-energy $e^+ e^-$ collider operating at 
$\sim 10.58$ GeV and $\sim 10.54$ GeV center-of-mass energies.
A fit to the Dalitz plot intensity distribution corresponding to the coherent superposition 
of amplitudes describing $\Lambda a_0(980)^+$ and $\Xi(1690)^{0} K^+$ production 
yields mass and width values of 
$1684.7{\pm 1.3}\;(\rm{stat.})\,^{+2.2}_{-1.6}\;(\rm{syst.})\;\; \rm{MeV}/c^2, $ and 
$8.1_{-3.5}^{+3.9}\;(\rm{stat.})\,^{+1.0}_{-0.9}\;(\rm{syst.})\;\; \rm{MeV}, $
respectively, for the $\Xi(1690)^0$,
while the spin is found to be consistent with value of 1/2 on the basis of studies 
of the $(\Lambda K_S)$ angular distribution.

\vfill
\begin{center}

Submitted to the 33$^{\rm rd}$ International Conference on High-Energy Physics,
ICHEP 06,\\
26 July---2 August 2006, Moscow, Russia.

\end{center}

\vspace{1.0cm}
\begin{center}
{\em Stanford Linear Accelerator Center, Stanford University,
Stanford, CA 94309} \\ \vspace{0.1cm}\hrule\vspace{0.1cm}
Work supported in part by Department of Energy contract DE-AC03-76SF00515.
\end{center}

\newpage
} 

\begin{center}
\small

The \babar\ Collaboration,
\bigskip

%
{B.~Aubert,}
{R.~Barate,}
{M.~Bona,}
{D.~Boutigny,}
{F.~Couderc,}
{Y.~Karyotakis,}
{J.~P.~Lees,}
{V.~Poireau,}
{V.~Tisserand,}
{A.~Zghiche}
\inst{Laboratoire de Physique des Particules, IN2P3/CNRS et Universit\'e de Savoie,
 F-74941 Annecy-Le-Vieux, France }
% ...

\end{center}\newpage

\section{INTRODUCTION}

Although considerable advances have been made in baryon spectroscopy over the past decade, there has been very
little improvement in our knowledge of hyperon resonances since 1988.
In particular, little is known about cascade resonances.
Aside from the $\Xi(1530)$, all $\Xi$ resonances rate below four stars according to the PDG ranking criteria~\cite{ref:pdg}.
The $\Xi(1690)$ has been observed in the $\Lambda \bar{K}$, $\Sigma \bar{K}$ and $\Xi \pi$
final states with various degrees of certainty.  However, its quantum numbers have not yet been measured.

The first evidence for the $\Xi(1690)$ came from the observation of a threshold enhancement in
the $\Sigma^{+, 0} K^{-}$ mass spectrum, produced in the reaction $K^- p\rightarrow (\Sigma^{+, 0} K^{-}) K \pi$ at 4.2 GeV/$c$
in a bubble chamber experiment~\cite{ref:r2}.  There were also indications of signals in the $(\Lambda \bar K^0)$ and  $(\Lambda K^-)$ channels.  
Subsequently, the $\Xi(1690)^-$ was observed in a hyperon beam experiment at CERN,
in which an enhancement around 1700 MeV$/c^2$ was seen in $(\Lambda K^-)$ pairs diffractively produced by
a 116 GeV/$c$ $\Xi^-$ beam~\cite{ref:r3,ref:r4}.   
The $\Xi^- \pi^+$ decay mode of the $\Xi(1690)^0$ was first reported by the WA89 Collaboration 
on the basis of a clear peak in the $\Xi^- \pi^+$ mass spectrum resulting 
from the interactions of a 345 GeV/$c$ $\Sigma^-$ beam in copper and carbon targets~\cite{ref:r5}.  
Evidence of $\Xi(1690)^0$ production in $\Lambda_c^+$ decay was reported by the Belle experiment, on the basis of $246\pm 20$
$\Lambda_c^+ \rightarrow (\Sigma^+ K^-) K^+$ and $363\pm 26$ $\Lambda_c^+ \rightarrow (\Lambda \bar{K}^0) K^+$ events~\cite{ref:Bel}.

\section{EVENT SELECTION}

In this paper, measurements of the mass and width of 
the $\Xi(1690)^0$ are presented.  The spin of the $\Xi(1690)^0$ is also investigated.  
The $\Xi(1690)^0$ is observed in the $\Lambda \bar K^0$ 
system produced in  
the decay $\Lambda_c^+ \rightarrow (\Lambda \bar K^{0}) K^{+}$, where the 
$\bar K^{0}$ is reconstructed as a $K_S\rightarrow \pi^+ \pi^-$~\cite{ref:cc}.  The data sample, 
collected with the \babar\ detector
at the \pep2\ asymmetric-energy $e^+e^-$ collider at center-of-mass energies $\sim 10.58$ and $\sim 10.54$ GeV, 
corresponds to a total integrated luminosity of about 200 fb$^{-1}$.
The detector is described
in detail elsewhere~\cite{ref:babar}.

The selection of $\Lambda_{c}^{+}$ candidates requires
the intermediate reconstruction of oppositely-charged track pairs consistent with
$\Lambda \rightarrow p \: \pi^-$ and 
$K_S \rightarrow \pi^+ \: \pi^-$ decays.    
The $\Lambda$ candidate is required to have 
invariant mass within a $\pm 3\sigma$ mass window centered on its peak position,
where $\sigma$ is the mass resolution obtained from a fit to the invariant mass spectum.
The selected $K_S$ invariant mass is within $\pm 25$ MeV$/c^2$ of the nominal value~\cite{ref:pdg}. 
Each intermediate state invariant mass is then constrained to its nominal value~\cite{ref:pdg}, 
with corresponding vertex fit probability required to be greater than 0.001.
The $\Lambda$ and $K_S$ candidates are then vertexed with a positively charged 
kaon track to form a $\Lambda_c^+$ candidate.  
In the reconstruction, proton and kaon candidates are required to satisfy identification criteria 
based on specific energy loss (${\rm d}E/{\rm d}x$) and Cherenkov angle measurements~\cite{ref:babar}.  
Since the $\Lambda$ and $K_S$ have displaced vertices, 
the signal-to-background ratio is improved by requiring that each  
decay vertex be separated 
from the event primary vertex 
by at least of $2.5$~mm in the flight direction of the 
$\Lambda$ or $K_S$. 
In order to further enhance the signal-to-background ratio the 
center-of-mass momentum $p^*$ of the $\Lambda_c^+$ is required to be greater than $0.5$ GeV$/c$.

The invariant mass spectrum of $\Lambda_{c}^{+}$ candidates satisfying 
these selection criteria is shown
before efficiency-correction in Fig.~1.
A signal yield of $2748\pm 297$ candidates is obtained
from a fit to the invariant mass spectrum with a double Gaussian 
signal function and a linear background. 
 
\begin{figure}[!t]
  \centering\small
  \includegraphics[height=7cm]{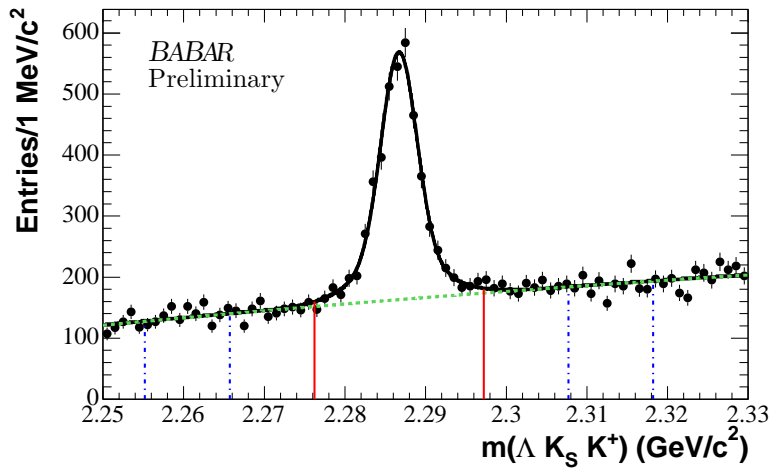} 
\begin{picture}(0.,0.)
    \put(-245,160){\normalsize\babar}
    \put(-245,150){Preliminary}
    \end{picture}
  \caption{The invariant mass distribution of uncorrected $\Lambda K_{S} K^{+}$ candidates  
in $\sim$ 200 fb$^{-1}$ of data. 
The superimposed curve corresponds to a binned $\chi^2$ fit which uses a double Gaussian signal function and 
a linear background  parametrization denoted by the dashed line.
The vertical lines delimit the signal region used in this analysis (solid) and the 
corresponding mass-sideband regions (dotted). }
 \label{fig:MassPlot}
\end{figure}

\section{DALITZ PLOT FOR $\Lambda_c^+\rightarrow \Lambda \bar K^{0} K^{+}$}

The Dalitz plot of $\Lambda_c^+\rightarrow \Lambda \bar K^{0} K^{+}$ signal candidates is 
shown, without efficiency-correction, in Fig.~2(a).  A clear band is observed in the 
mass-squared region of the $\Xi(1690)^0$, together with an accumulation of events in the $\bar K^0 K^+$ 
threshold region.   
In contrast, the Dalitz plots corresponding to the $\Lambda_c^+$ mass-sideband regions exhibit no structure.  
The $\Lambda_c^+$ mass-sideband-subtracted $\Lambda K_{S}$ invariant mass projection without efficiency-correction 
is shown in Fig.~2(b).  
A clear signal for the $\Xi(1690)^0$ resonance is observed.  

\begin{figure}[!t]
  \centering\small
  \includegraphics[height=6cm]{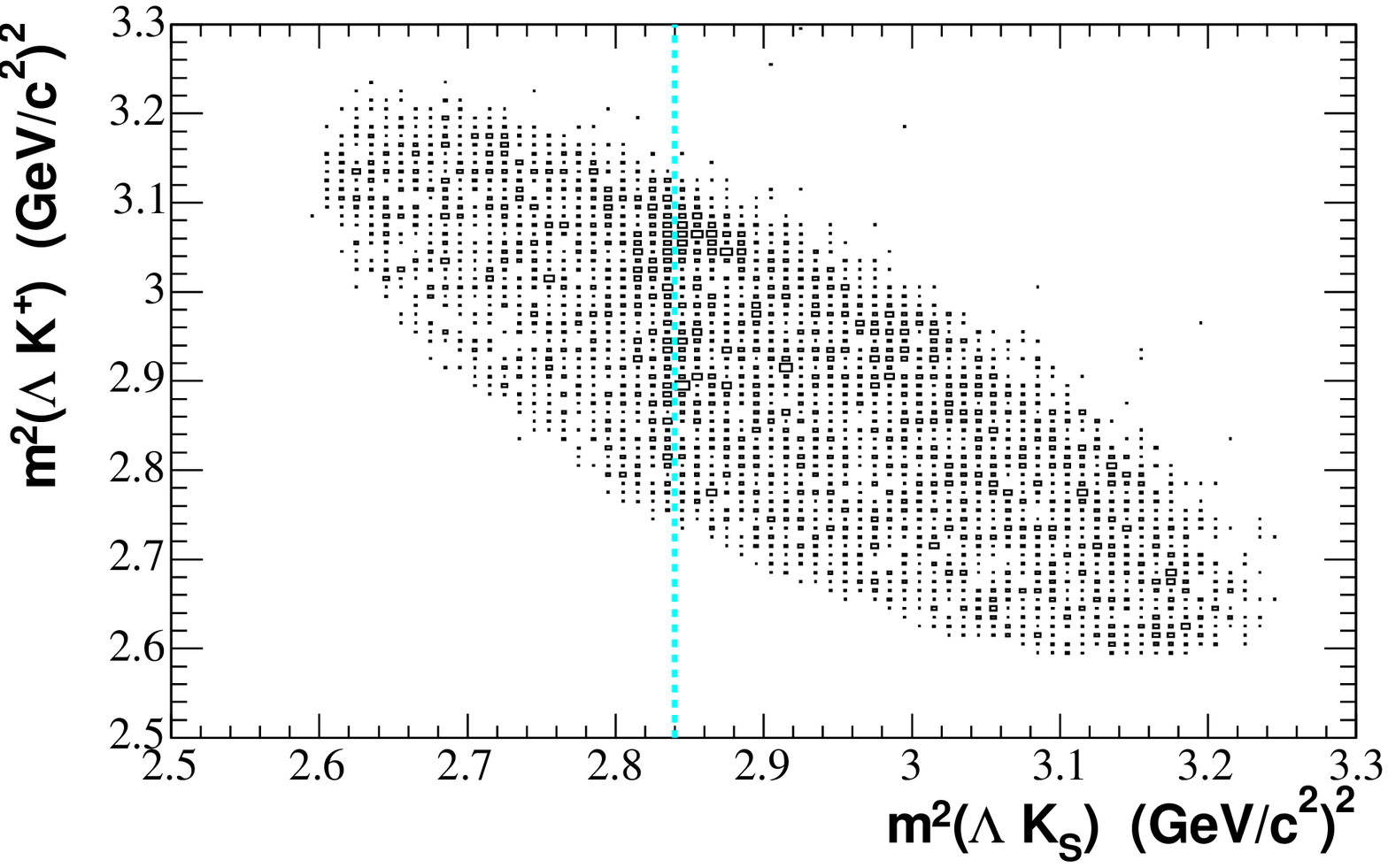} 
  \includegraphics[height=6cm]{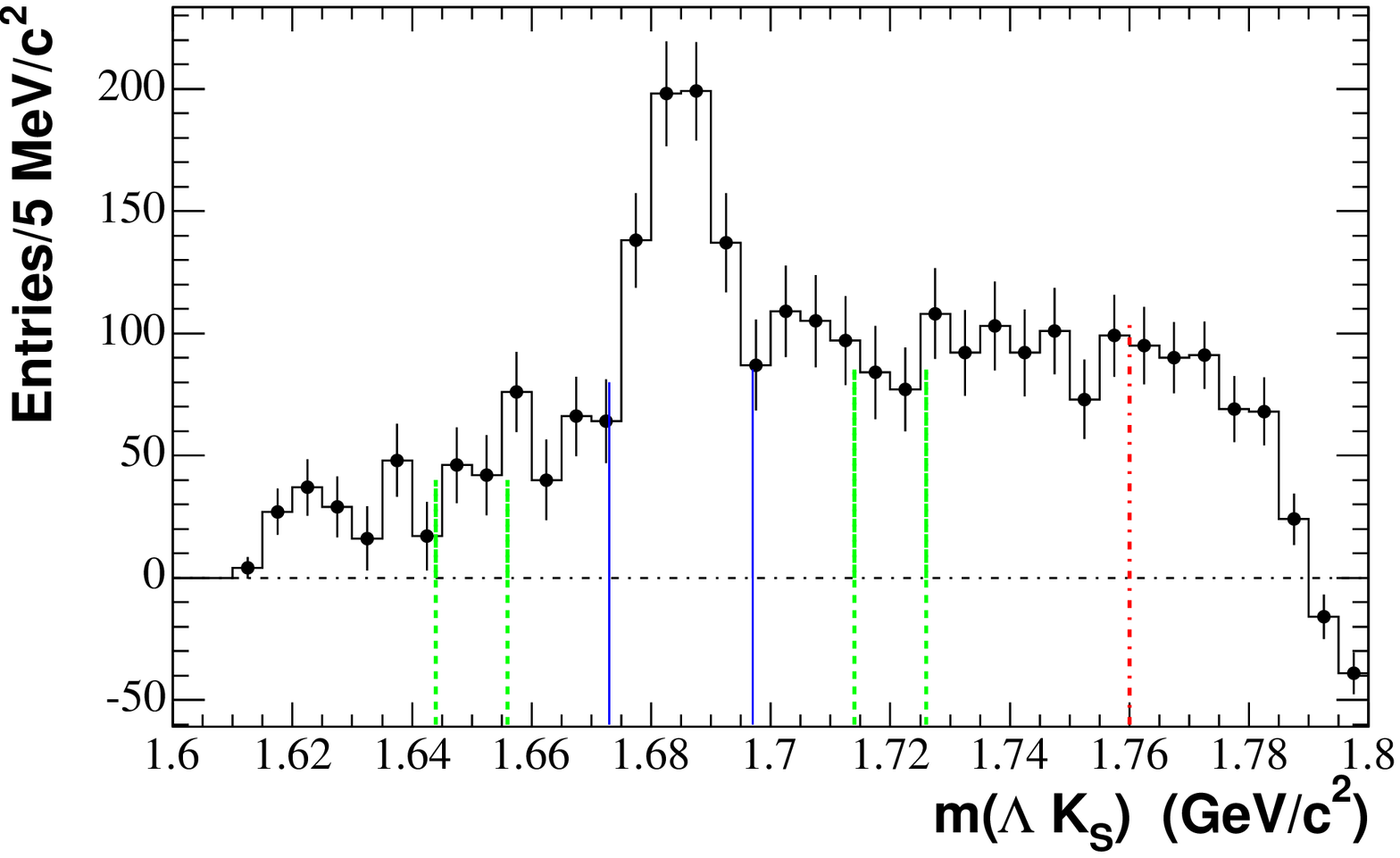}
\begin{picture}(0.,0.)
    \put(-75,312){\normalsize\babar}
    \put(-75,302){Preliminary}
    \put(-75,140){\normalsize\babar}
    \put(-75,130){Preliminary}
    \put(-215,310){\bf{(a)}}
    \put(-215,140){\bf{(b)}}
    \end{picture}
  \caption{(a) The Dalitz plot for $\Lambda_c^+\rightarrow \Lambda \bar K^0 K^{+}$ 
corresponding to the $\Lambda_c^+$ signal region indicated in Fig.~\ref{fig:MassPlot}.  
The dashed blue line indicates the nominal mass-squared region of the $\Xi(1690)^0$.    
(b) The $\Lambda_c^+$ mass-sideband-subtracted $\Lambda K_{S}$ invariant mass projection.  
The solid lines delimit the
$\Xi(1690)^0$ mass-signal region, while the dotted lines indicate the mass-sideband regions.  
The dashed line at $1.76$ GeV/$c^2$ indicates the upper limit of the 
uniformly accessible mass range, and corresponds to the lower bound of the
$\Lambda_c^+$ low-mass sideband.}
 \label{fig:DalitzPlot}
\end{figure}

\section{EFFICIENCY PARAMETRIZATION}

The selection efficiency is determined from a sample of $\Lambda_c^+ \rightarrow \Lambda \bar K^{0} K^+$ Monte Carlo 
events uniformly distributed across phase-space on the Dalitz plot, and is parametrized over the entire Dalitz plot.        
For the measurement of the mass and width of the $\Xi(1690)^0$, the $(\Lambda K_S)$ invariant mass spectrum is 
corrected according to an efficiency parametrized in two dimensions as a function of the 
cosine of the angle of the $\Lambda$ in the $(\Lambda K_S)$ rest-frame with respect to the $(\Lambda K_S)$ 3-momentum 
in the $\Lambda_c^+$ rest-frame (i.e. the $\Lambda$ helicity angle),  cos$\theta_\Lambda$, and of $m(\Lambda K_S)$.  
Each selected event is weighted inversely according to this parametrization.  

\section{MASS AND WIDTH MEASUREMENT}

\subsection{Resolution Smearing}

For a narrow resonance such as the $\Xi(1690)$, the measurement of its mass and 
width is sensitive to detector resolution effects;  
in particular the apparent width will be larger than its true value.  
Therefore, in this measurement, the fit function is smeared by a mass resolution function, obtained 
from the Monte Carlo phase-space Dalitz plot in the region of the $\Xi(1690)^0$, 
and consisting of two Gaussians with a common center; a narrow Gaussian with $\sigma = 1.07 \pm 0.02$ MeV/$c^2$ and 
a broader Gaussian with $\sigma = 2.44 \pm 0.06$ MeV/$c^2$.  The narrow Gaussian represents 70\% of the lineshape.  

In order to determine the mass and width parameters of the $\Xi(1690)^0$ resonance 
we parametrize its $\Lambda_c^+$ mass-sideband-subtracted background distribution from a model of the Dalitz plot intensity distribution 
containing terms describing a coherent superposition of two amplitudes.  The interference 
between these has a significant impact on the interpretation of the apparent $\Xi(1690)^0$ 
signal.  The model employed and the results obtained from it are discussed in the following sub-sections.

\subsection{Study of the $\Xi(1690)^0$ background lineshape}

The accumulation of events near the $\Xi(1690)^0$ band, towards the upper boundary of the Dalitz plot shown in Fig.~2(a),
 is consistent with the 
coherent superposition of amplitudes characterizing 
 $(\Lambda a_0(980)^+)$ and $(\Xi(1690)^{0} K^+)$ decay of the $\Lambda_c^+$.  

The $a_0(980)$ is known to couple to both $\eta \pi$ and $\bar K K$ and is therefore characterized by the following
 Flatt\'{e} parametrization~\cite{ref:Flatte}:
\begin{eqnarray}A(a_0[980])=\frac{g_{\bar{K} K}}{m^{2}_{0}-m^{2}-i(\rho_{\eta \pi}g^2_{\eta \pi}+\rho_{\bar{K} K}g^2_{\bar{K} K})},\end{eqnarray}
where $\rho_j(m)=2q_j/m$ is the phase space factor for the decay into the two-body channel $j=\eta \pi, \bar{K} K$.
The coupling constants measured by the Crystal Barrel~\cite{ref:crystalbar}
 and \babar~\cite{ref:antimo} experiments, respectively are:
$$g_{\eta \pi}=324\pm 15\; \rm{MeV}$$ and $$g_{\bar{K} K}=646\pm 29\; \rm{MeV}.$$
The pole value obtained by the Crystal Barrel experiment is:
$$m_0=999\pm 2\; \rm{MeV}/c^2.$$

It is assumed that the $a_0(980)^+$ is produced in an S-wave orbital angular momentum state with respect to the 
recoil $\Lambda$ (although P-wave is also allowed) and that as a result no additional form factor describing the $\Lambda_c^+$ 
decay to the $\Lambda a_0(980)^+$ final state is required.  It then 
follows that Eq.~(1) describes the amplitude for this decay mode at the $\Lambda_c^+$, 
with isotropic decay of the $a_0(980)^+$ to $\bar K^0 K^+$ implied.  

The angular distribution within the $\Xi(1690)^0$ band in the Dalitz plot (Fig.~2(a)) is consistent with being flat 
(see section on Spin Study).  
Consequently, the amplitude describing the $\Xi(1690)^0$ is chosen to be 
\begin{eqnarray} A(\Xi[1690]) = \frac{1}{(m_0^2-m^2)-i m_0\Gamma(m)} ,\end{eqnarray}
where
$m_0=m(\Xi(1690)^0)$.  
Assuming that the $\Lambda K_S$ system is in an S-wave state, and ignoring the contribution of other 
partial widths to the total width, the latter is described by                                                                               
\begin{eqnarray} \Gamma(m)=\Gamma(m_0)\frac{q}{m}\frac{m_0}{q_0}, \end{eqnarray} where $\Gamma(m_0)$ is the total 
width parameter to be extracted from the data, and $q$ ($q_0=q(m_0)$) is the momentum of the $\Lambda$ in the $(\Lambda K_S)$ rest-frame.  
As for the $\Lambda a_0(980)^+$ amplitude, it is assumed that the $\Xi(1690)^0$ is produced in an orbital angular momentum S-wave 
with respect to the recoil $K^+$ (although P-wave is also allowed), and that no additional form factor describing $\Lambda_c^+$ 
decay to $\Xi(1690)^0 K^+$ is required.  It then follows that Eq.~(2) describes this decay amplitude of the $\Lambda_c^+$ 
with isotropic decay to $\Lambda \bar K^0$ implied.

The model is then used to describe the intensity distribution at a point on the Dalitz plot by means of the squared modulus 
of a coherent superposition of these two amplitudes as follows:
\begin{eqnarray} |A|^2 = |r e^{i\phi}A(\Xi[1690])+A(a_0[980])|^2 ,\end{eqnarray}
where $r$ represents a constant relative strength of the amplitudes and $\phi$ is a constant relative phase between them.

\begin{figure}[!b]
  \centering\small
  \includegraphics[width=.5\textwidth]{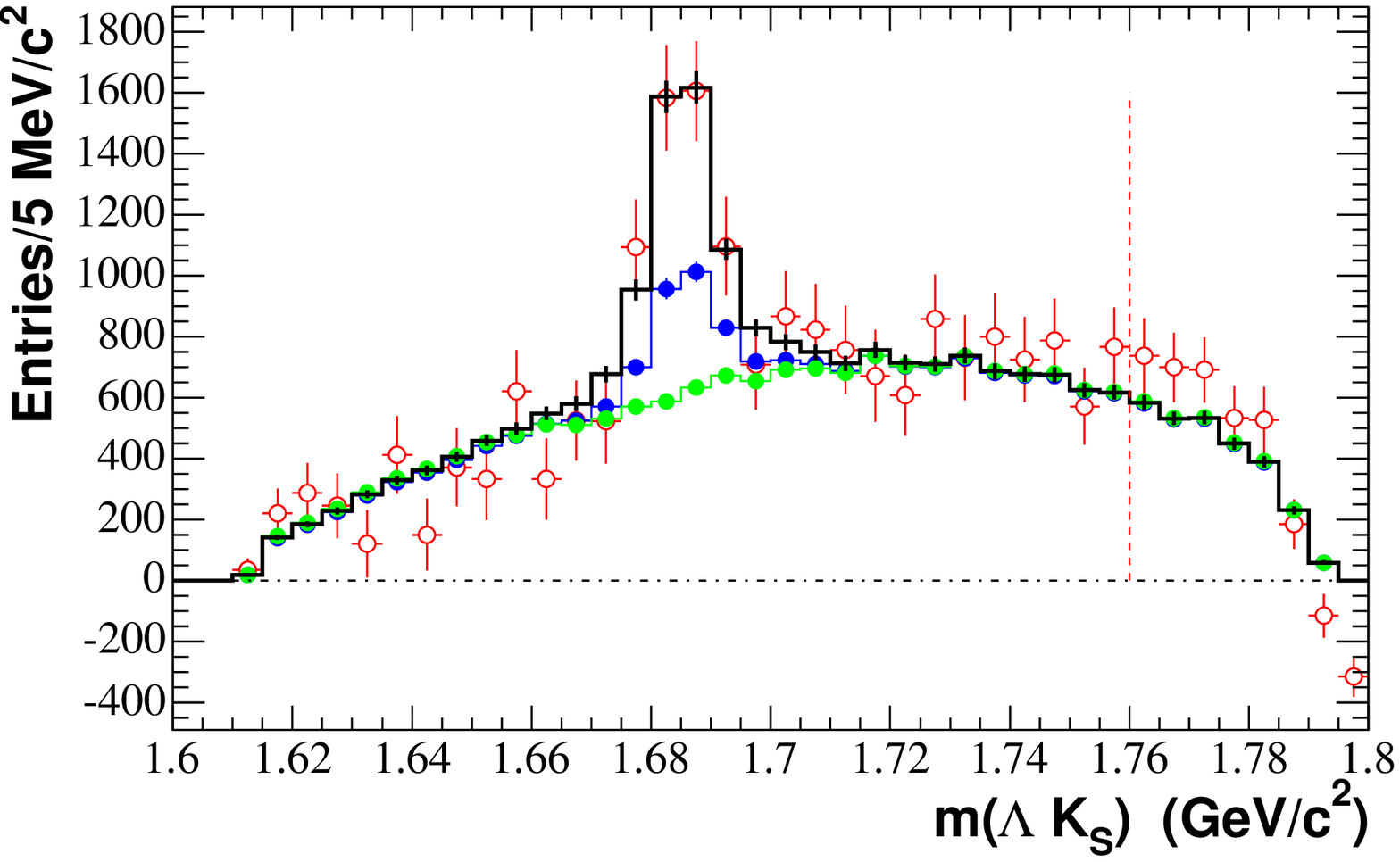}
  \includegraphics[width=.5\textwidth]{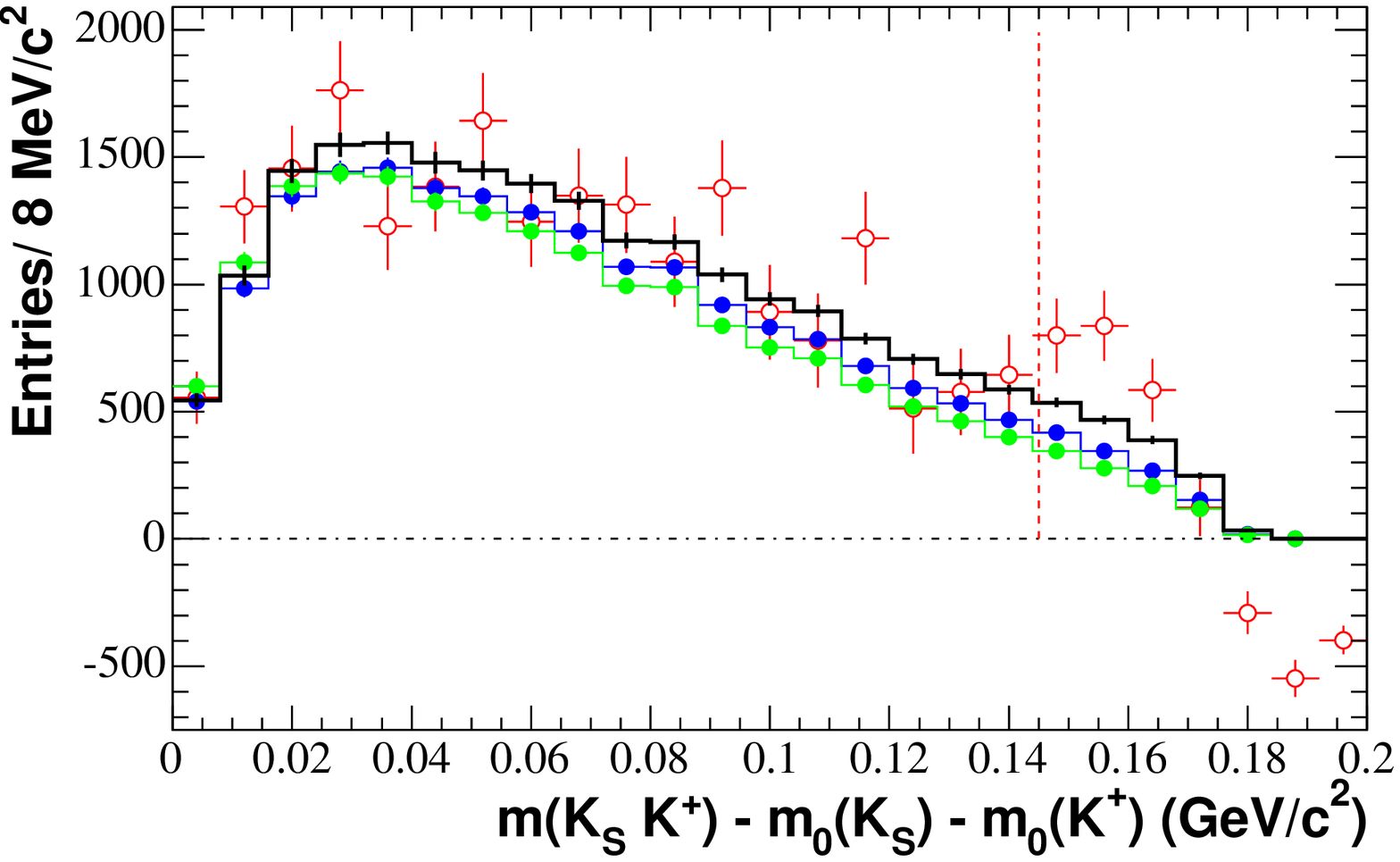}
  \includegraphics[width=.5\textwidth]{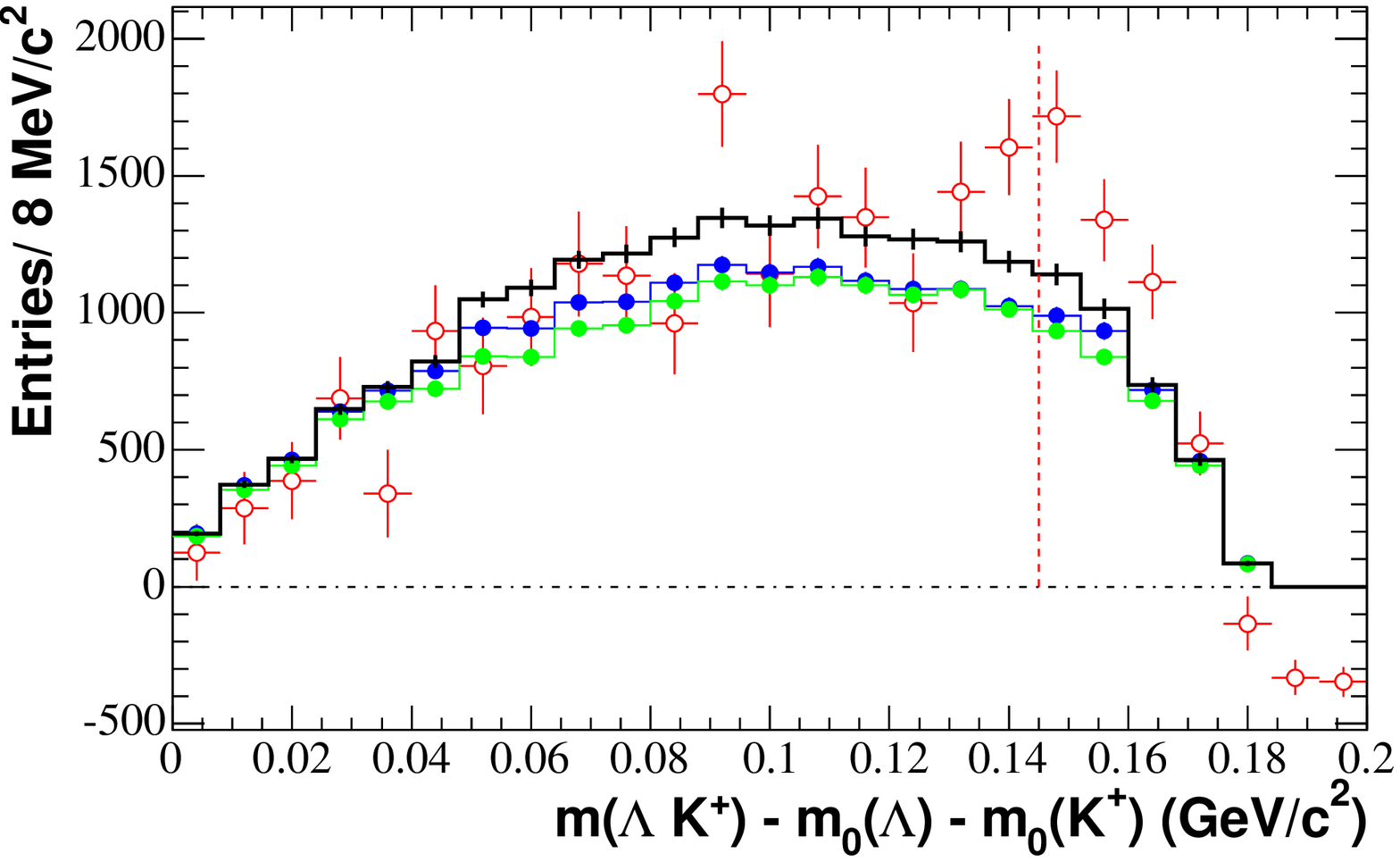}
  \begin{picture}(0.,0.)
    \put(-70,450){\normalsize\babar}
    \put(-70,440){Preliminary}
    \put(-200,450){\bf{(a)}}
    \put(-200,290){\bf{(b)}}
    \put(-200,130){\bf{(c)}}
    \end{picture}
 \caption{The $m(\Lambda K_S)$ (a), $(K_S K^+)$ mass difference with respect to threshold (b), 
and $(\Lambda K^+)$ mass difference with respect to threshold (c) Monte Carlo projections weighted by $|A|^2$ 
are shown as the histograms superimposed on the open circles corresponding to the efficiency-corrected distributions in data.  
The light-colored (green) and dark-colored (blue) points correspond to the Monte Carlo projections weighted by the terms 
$|A(a_0[980])|^2$ and $|A|^2-r^2|A(\Xi[1690])|^2$, respectively, where these model parameters are given in Eq.~4. As in Fig.~2(b), 
the vertical dashed lines indicate the upper limit of the uniformly accessible mass range. }
  \label{fig:Par34}
\end{figure}

In order to test the model and obtain appropriate values for $r$ and $\phi$, 
a width $\Gamma(m_0)=10$ MeV is assumed for the $\Xi(1690)^0$ in the expression for $A(\Xi[1690])$.  
The corresponding values for $r$, and $\phi$ are then obtained from a simultaneous $\chi^2$ minimization between the simulated 
$\Lambda K_S$, $K_S K^+$ and $\Lambda K^+$ invariant mass spectra and the corresponding projections in data.  
The minimum $\chi^2$ corresponds to $\phi = 34$ deg. and $r=0.025$.  
In Fig.~3, the $m(\Lambda K_S)$, $m(K_S K^+)$ and $m(\Lambda K^+)$ Monte Carlo projections weighted by the resulting $|A|^2$ 
are represented by the histograms superimposed on the open circles corresponding to the efficiency-corrected distributions in data, 
and it is clear that the simple model describes the mass projections very well.  
The green and blue points correspond to the Monte Carlo projections weighted by the terms 
$|A(a_0[980])|^2$ and $|A|^2-r^2|A(\Xi[1690])|^2$, 
respectively, where the latter represents the resulting total background in the $(\Lambda K_S)$ mass distribution (after 
$\Lambda_c^+$ mass-sideband-subtraction).  
The main feature of this background is the presence of a peak in the $\Xi(1690)^0$ signal region that results from the interference 
between the $\Xi(1690)^0$ and the $a_0(980)^+$ amplitudes.

\subsection{Measurement Procedure}

Previous measurements of the mass and width of the $\Xi(1690)$ resonance relied on a fit to the
signal making an ad hoc polynomial assumption about the shape of an incoherent background~\cite{ref:Bel}.
Based on the agreement between the weighted Dalitz plot projections obtained from simulation
and the data of Fig.~3, it is apparent that a background description of the $m(\Lambda \bar{K}^0)$ 
spectrum based on a polynomial parametrization does not take adequate 
account of the physics process involved in the description of the Dalitz plot.  

The procedure adopted to fit the sideband-subtracted, efficiency-corrected $(\Lambda {K}_S)$ 
invariant mass distribution shown by the open circles of Fig.~4 is therefore as follows: 

\noindent (i)  The mass distribution is parametrized by means of the function   
\begin{eqnarray}
f = c_1\frac{pq}{M}|A(1690)|^2+c_2\left [ |A(980)|^2 + 2 Re(r e^{-i\phi}A(1690)^*A(980))\right ]. 
\end{eqnarray}
The first term represents the lineshape of the $\Xi(1690)^0$, and incorporates the 
resolution smearing procedure discussed previously.  The second term is the total 
background due to the $a_0(980)^+$ and its interference with the $\Xi(1690)^0$;  
$M$ is the $\Lambda_c^+$ mass, $p$ is the momentum of the $K^+$ in the $\Lambda_c^+$ rest-frame, 
and $q$ is the momentum of the $\Lambda$ in the $(\Lambda K_S)$ rest-frame.
The lineshape for the latter term is obtained by simulation, incorporating fixed values for $\phi$, 
$m_0$, and $\Gamma(m_0)$, obtained by the iteration described below.  The first term corresponds to the actual fit function 
where $m_0$ and $\Gamma(m_0)$ are parameters.

\noindent (ii) In the second term, the phase $\phi$ is fixed at values which are varied in one degree steps 
from $15\deg - 50\deg$.
The value for $r$ is obtained by the simultaneous minimization of the $\chi^2$ of the mass projections of Fig.~3.  

\noindent (iii)  The fit procedure for each choice of $\phi$ is iterative, in that the values of $r$, $m_0$ and $\Gamma(m_0)$ used in 
the second term are fixed to those from the previous iteration, and the value of $\Gamma(m_0)$ also incorporates 
the effect of resolution smearing; the $(\Lambda {K}_S)$ mass projection due to this term is obtained by weighting the MC phase
space distribution.  

\noindent (iv)  The procedure converges typically after two iterations in the sense that  $m_0$ and $\Gamma(m_0)$ 
no longer change. 

The width value of 10 MeV shown in Fig.~3 corresponds to the initial value of the iteration process.
The results of the best fit obtained by this procedure are shown in Fig.~4 in comparison to the 
background-subtracted, efficiency-corrected $(\Lambda {K}_S)$ mass distribution (open circles with error bars).  

The value $\Gamma(m_0)=9.5$ MeV with $m_0=1684.7$ MeV/$c^2$ used in the second term of Eq.~(5) was obtained by 
the iteration procedure, starting at $\Gamma(m_0)=10$ MeV with $m_0=1685.0$ MeV/$c^2$.  
The histogram represents the total contribution  in each mass interval obtained from the fit.

The value for the total $\Xi(1690)^0$ signal yield obtained from this fit is $1837\pm 463$ events.  
The peaking background due to the interference with
the $a_0(980)^+$ results in a signal yield about 30\% smaller than would be obtained with a fit using an 
incoherent polynomial background.

\begin{figure}[!b]
  \centering\small
  \includegraphics[height=7cm]{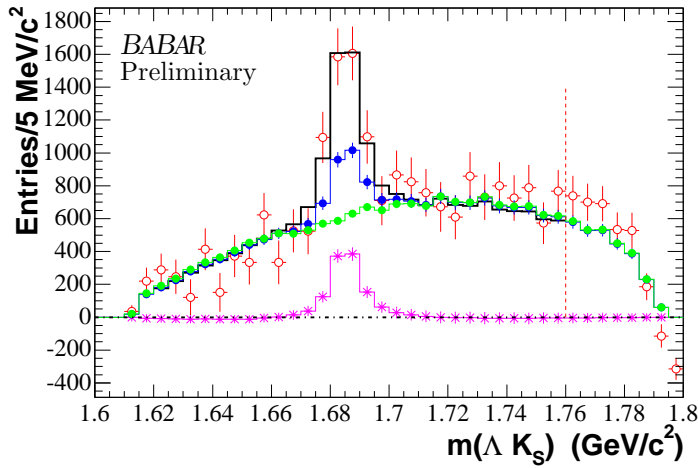}
\begin{picture}(0.,0.)
    \put(-230,162){\normalsize\babar}
    \put(-230,152){Preliminary}
    \end{picture}
  \caption{The $(\Lambda K_S)$ invariant mass projection in data, after all corrections, is represented by the open circles.
Superimposed is the distribution for the simulated interference term
 $2Re(r e^{-i\phi}A(1690)^*A(980))$ (represented by the stars), the $|A(980)|^2$ distribution (light-colored dots), 
and (dark-colored dots) the total background obtained as the sum of these  
contributions.  The value $\Gamma(m_0)=9.5$ MeV, after resolution smearing, (obtained by 
iteration), and $\phi=34$ deg. have been used in obtaining the background distribution.  
The histogram represents the total contribution in each mass interval, and is obtained from the fit as the 
sum of the $\Xi(1690)^0$ signal smeared by resolution and the total background contribution. This fit has a $\chi^2/NDF$ of 20/24.}
 \label{fig:FitPar2Plot}
\end{figure} 

\clearpage
\subsection{Systematic Uncertainties}

The dominant sources of systematic uncertainty in this measurement are due to the choice of phase, $\phi$,
entering the $\Xi(1690)^0$ background distribution through the second term in Eq.~(5).   
From the fits described above for different choices of phase value, it is found that the values yielding 
a $\chi^2$ increase of one unit relative to the minimum are $\phi=21\deg$ and $\phi=46\deg$, respectively. 
The shape of the interference term between the $a_0(980)^+$ and the $\Xi(1690)^0$ amplitudes 
has a tail to high mass and a sharp rise at low mass for $\phi=21\deg$ and vice-versa for $\phi=46\deg$.
This effect causes the fitted $\Xi(1690)^0$ mass value to shift significantly, and this must be taken into account as a
systematic uncertainty.  The corresponding effect on the width proves to be quite small.

A second source of systematic uncertainty results from the choice of $\Lambda_c^+$ mass sidebands.  
Up to this point, the $\Lambda_c^+$ mass sidebands used for the $\Xi(1690)^0$ mass and width measurement as 
well as for the study of systematic uncertainties corresponded to $\pm 6\sigma$ to $\pm 9\sigma$ from the $\Lambda_c^+$ fitted mean.
The entire fit procedure is repeated first using
 $\Lambda_c^+$ mass sidebands which are from $\pm 7.5\sigma$ to $\pm 10.5\sigma$ from the $\Lambda_c^+$ fitted mean, 
and then $\Lambda_c^+$ mass sidebands which 
are from $\pm 4.5\sigma$ to $\pm 7.5\sigma$ from the $\Lambda_c^+$ fitted peak value.
The associated systematic uncertainties in mass and width are then estimated as one half of the difference in the 
fit values obtained.

The effect of the efficiency parametrization on the fit results is found to be negligible.

In the definition of the $\Xi(1690)^0$ amplitude used in the fits it has been 
assumed that the orbital angular momentum in $\Lambda_c^+$ decay is $L=0$ and that the 
orbital angular momentum in the decay of the $\Xi(1690)^0$ to $\Lambda \bar K^0$ system is $l=0$.
The fit procedure is repeated using $L=0$ and $l=1$,
$L=1$ and $l=1$, and $L=1$ and $l=0$.  
The effect on fitted mass and width is small, and corresponding systematic uncertainties 
are estimated as the extrema of the changes which result.

Effects due to the uncertainties in the respective widths of the narrow and 
wide Gaussians used to parametrize the lineshape of
 the resolution function are also considered.  
The fitting procedure is repeated incorporating resolution smearing,
with the values of the Gaussian widths shifted by $\pm 1$ standard deviation, 
and the resulting mass and width changes are found to be small.  

The width of the $\Lambda_c^+$ signal in data and MC agree to better than $10\%$.  
The width of the resolution function used in smearing the $\Xi(1690)^0$ lineshape is therefore increased by $10\%$, 
and the fitted value of the $\Xi(1690)^0$ width is found to decrease by 0.2 MeV.  This effect is also included in the 
estimate of systematic uncertainty associated with resolution.

Uncertainties due to detector effects are estimated from the study of
 systematic uncertainties in the mass measurement of the $\Lambda_c^+$
using $\Lambda_c^+ \rightarrow \Lambda K_S K^+$ and $\Lambda_c^+ \rightarrow \Sigma K_S K^+$ decays~\cite{ref:brian}.
This study found that the dominant systematic uncertainty in mass
 arose from the amount of material in the tracking volume and from the
magnetic field strength, but that this effect was small for $\Lambda_c^+ \rightarrow \Lambda \bar K^0 K^+$ because of the 
limited phase space available in the decay.
Because systematic uncertainties scale with Q-value, we obtain a conservative estimate of the uncertainty due to 
detector effects in the decay $\Lambda_c^+ \rightarrow \Xi(1690)^0 K^+$ based
 the uncertainty determined from $\Lambda_c^+ \rightarrow \Lambda \bar K^0 K^+$.
Thus, we quote a systematic uncertainty of $\pm 0.1$  MeV/$c^2$ on the $\Xi(1690)^0$ mass.

\begin{table}[t!]
\begin{center}
\begin{tabular}{ l r r }
\hline \hline
Source of Syst. Error & Mass Error & Width Error \rule{0cm}{0.5cm} \\
\hline
choice of relative phase &$^{+2.1}_{-1.5}$ MeV/$c^2$ & $+0.3$ MeV \rule{0cm}{0.5cm} \\
choice of $\Lambda_c^+$ mass sidebands & $\pm 0.4$ MeV/$c^2$ & $\pm 0.8$ MeV \rule{0cm}{0.5cm} \\
efficiency parametrization & $\pm 0.1$ MeV/$c^2$ & $\pm 0.3$ MeV \rule{0cm}{0.5cm}\\
choice of $L$, $l$ & $\pm 0.2$ MeV/$c^2$ & $\pm 0.3$ MeV \rule{0cm}{0.5cm}\\
resolution function lineshape  & $-0.1$ MeV/$c^2$ & $^{+0.1}_{-0.2}$ MeV \rule{0cm}{0.5cm} \\
detector effects &  $\pm 0.1$  MeV/$c^2$ & 0.0 MeV\rule{0cm}{0.5cm}\\
\hline
Total Systematic Error & $^{+2.2}_{-1.6}$ MeV/$c^2$ & $^{+1.0}_{-0.9}$ MeV \rule{0cm}{0.5cm} \\
\hline \hline
\end{tabular}
\end{center}
 \caption{Summary of systematic uncertainties associated with the measurements of the mass and width of the $\Xi(1690)^0$.}
 \label{tab:syssum}
\end{table}
Table~\ref{tab:syssum} summarizes the systematic uncertainties associated with the mass and width measurements of the $\Xi(1690)^0$.

\bigskip
\bigskip
\bigskip
\subsection{Summary of Results}

The measured values of the mass and width of the $\Xi(1690)^0$ obtained in the present analysis are as follows:

$$m(\Xi(1690))= 1684.7{\pm 1.3}\;(\rm{stat.})\,^{+2.2}_{-1.6}\;(\rm{syst.})\;\; \rm{MeV}/c^2, $$
$$\Gamma(\Xi(1690))=8.1_{-3.5}^{+3.9}\;(\rm{stat.})\,^{+1.0}_{-0.9}\;(\rm{syst.})\;\; \rm{MeV}.$$

The uncertainty in the mass value is mainly systematic, and results primarily from the interference between the 
$\Xi(1690)^0$ and $a_0(980)^+$ amplitudes.  In contrast, the width uncertainty is primarily statistical in nature.

\section{SPIN STUDY}

For the decay of a spin 1/2 charm baryon
to a hyperon and a pseudo-scalar meson (where the former decays to a secondary hyperon and a pseudo-scalar meson),
the angular distribution of the decay products can be determined unambiguously,
by choosing the quantization axis along
the direction of the primary hyperon in the charm baryon rest-frame, such that
the helicity $\lambda_i$ of the primary hyperon can only take the values  $\pm 1/2$.
The probability for the secondary hyperon to emerge with Euler angles $(\phi, \theta, 0)$ with respect to the quantization axis,
is given by the square of the amplitude characterizing the decay of a primary hyperon (in this case the $\Xi(1690)^0$ resonance) with
total angular momentum $J$ and helicity $\lambda_{i}$
to a 2-body system with net helicity $\lambda_{f}$:
$$A^J_{\lambda_f} D^{J *}_{\lambda_{i} \lambda_{f}}(\phi, \theta, 0), $$
where the transition matrix element $A^J_{\lambda_f}$
represents the coupling of the $\Xi(1690)^0$ to the final state; $A^J_{\lambda_f}$ does not depend on $\lambda_i$ because of
rotational invariance~\cite{ref:form,ref:form2,ref:Eckart,ref:fermi}.
The helicity angle $\theta$, illustrated in Fig.~5, is defined as the angle between the direction of the secondary
hyperon in the primary hyperon rest-frame
and the direction of the primary hyperon in the
charm baryon rest-frame.
The angular distribution of the secondary hyperon is then given by the total intensity,
$$I\propto \sum_{\lambda_{i}, \lambda_{f}} \rho_{i}\left |A^J_{\lambda_f}D^{J *}_{\lambda_{i} \lambda_{f}}(\phi, \theta, 0)\right |^2$$
\noindent where $\rho_i$ ($i= \pm 1/2$) are the diagonal density matrix elements
inherited from the charm baryon, and the sum is over all initial and final helicity states.
Previous studies~\cite{ref:omesp} have indicated that in an inclusive environment the parent 
baryon diagonal density matrix elements are equally populated.
Thus, the helicity angular distribution contains no dependence on the $\Lambda_c^+$ density matrix elements.
Using the above expression, the angular distribution of the $\Lambda$ decay product of the $\Xi(1690)^0$
is computed according to different spin hypotheses for the $\Xi(1690)^0$:
\begin{eqnarray}
J_{\Xi(1690)}=1/2 &:& 1\\
J_{\Xi(1690)}=3/2 &:& \frac{1}{4}(1 + 3{\rm cos}^2\theta_{\Lambda}) \\
J_{\Xi(1690)}=5/2 &:& \frac{1}{4}(1-2{\rm cos}^2\theta_{\Lambda}+5{\rm cos}^4\theta_{\Lambda})
\end{eqnarray}

The above equations ignore the presence of a coherent amplitude describing the $K_S K^+$ 
structure in the $\Lambda_c^+$ Dalitz plot.  As discussed above, there is evidence of such structure.  Nevertheless,  
the decay angular distribution of the $\Xi(1690)^0$ is investigated by means of a sideband subtraction procedure  
and the MC model of the Dalitz plot structure is used to take account of systematic effects due to $a_0(980)^+$ interference.  

\begin{figure}[!t]
  \centering\small
    \includegraphics[width=0.45\textwidth]{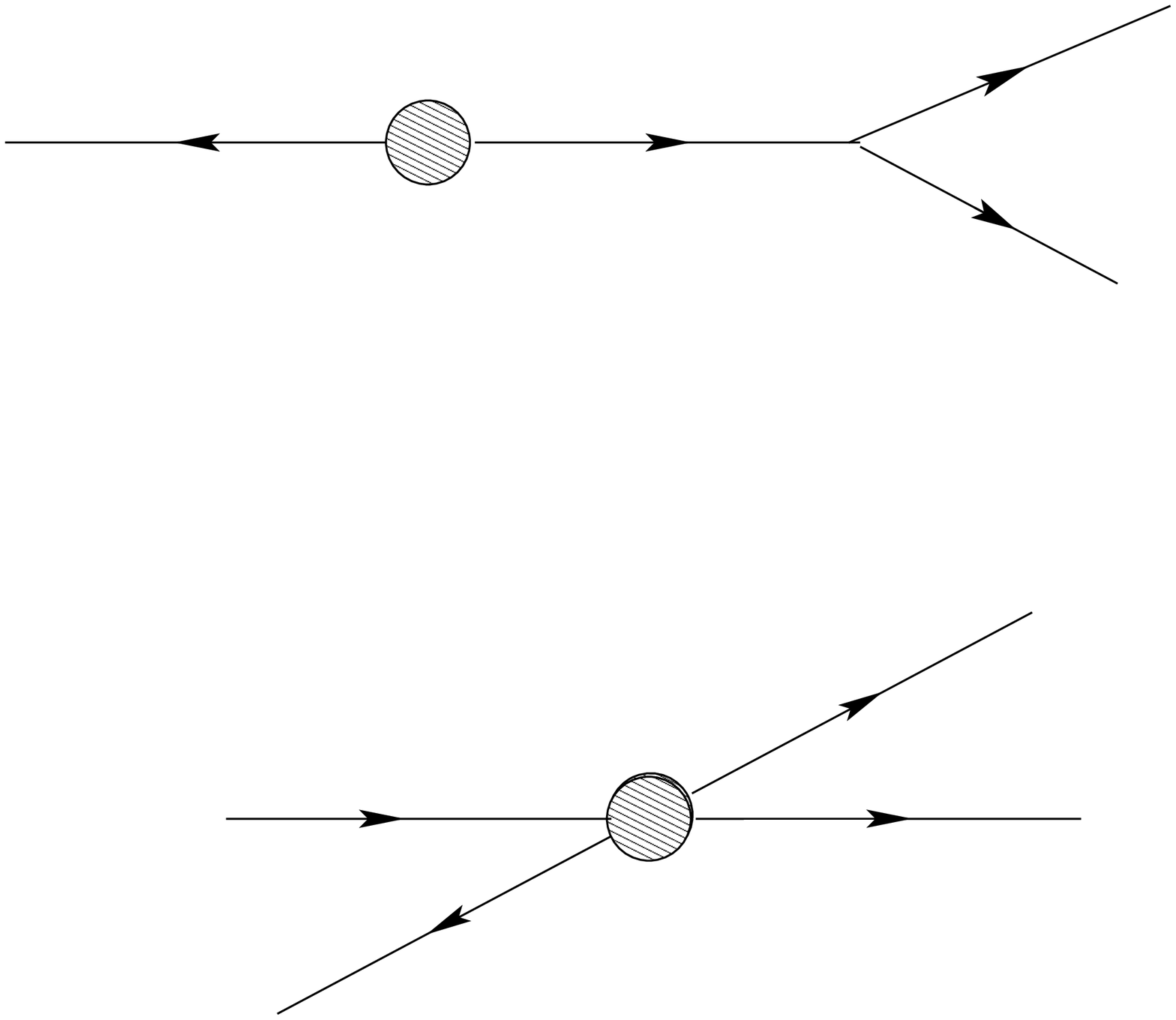}
    \begin{picture}(0.,0.)
    \put(-205,195){${K}^+_1$}
    \put(-155,190){${\Lambda}^+_c=\vec{0}$}
    \put(-90,195){${\Xi(1690)}_1$}
    \put(0,235){$\Lambda_1$}
    \put(-10,175){${K_S}_1$}
    \put(-210,160){a) All decay products in the $\Lambda^+_c$ rest-frame.}
    \put(-215,85){${\Xi(1690)}_1$}
    \put(-115,69){${\Xi(1690)}_2=\vec{0}$}
    \put(-80,92){)}
    \put(-68,93){\bf{$\theta_{\Lambda}$}}
    \put(-22,125){${\Lambda}_2$}
    \put(-178,45){${K_S}_2$}
    \put(-210,23){b) All decay products in the $\Xi(1690)^0$ rest-frame;}
    \put(-200,9){   in this frame, ${\Xi(1690)}_1 \rightarrow {\Xi(1690)}_2=\vec{0}$, }
    \put(-200,-2){   ${\Lambda}_1 \rightarrow {\Lambda}_2$, ${K_S}_1 \rightarrow {K_S}_2$.}
    \end{picture}
  \caption{Schematic definition of the helicity angle $\theta_{\Lambda}$ in the decay chain $\Lambda_c^+ \rightarrow \Xi(1690)^0 K^+$, $\Xi(1690)^0 \rightarrow \Lambda K_S$; as shown in b) $\theta_{\Lambda}$ is the angle between the $\Lambda$ direction in the $\Xi(1690)^{0}$ rest-frame and the $\Xi(1690)^0$ direction in the $\Lambda_c^+$ rest-frame.}
  \label{fig:Helicity}
\end{figure}

The $\Xi(1690)^0$ mass-sideband-subtracted cos$\theta_{\Lambda}$ distribution in data is shown in Fig.~6.
The distribution has been corrected with an efficiency calculated for each cos$\theta_{\Lambda}$ interval  
from phase-space $\Lambda_c^+\rightarrow \Lambda \bar K^0 K^+$ Monte Carlo. 
The horizontal line represents the expected distribution for $J_{\Xi(1690)}=1/2$,
the dashed curve corresponds to $J_{\Xi(1690)}=3/2$,
while the solid curve corresponds to $J_{\Xi(1690)}=5/2$ (Eqs.~(6) - (8), respectively). 
The fit C.L. values are summarized in Table~2.

\begin{figure}[!t]
  \centering\small
  \includegraphics[height=6cm]{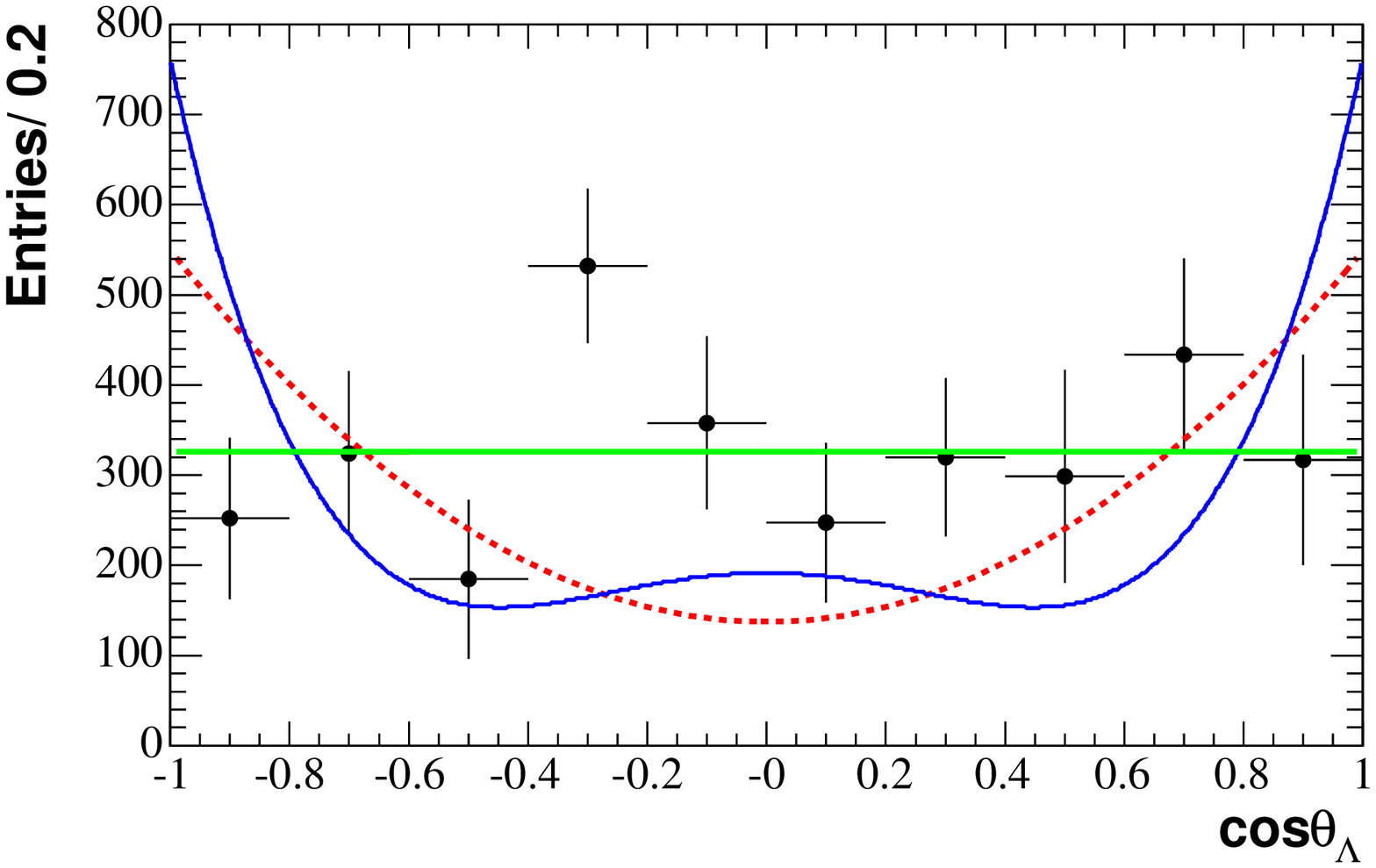}
  \includegraphics[height=6cm]{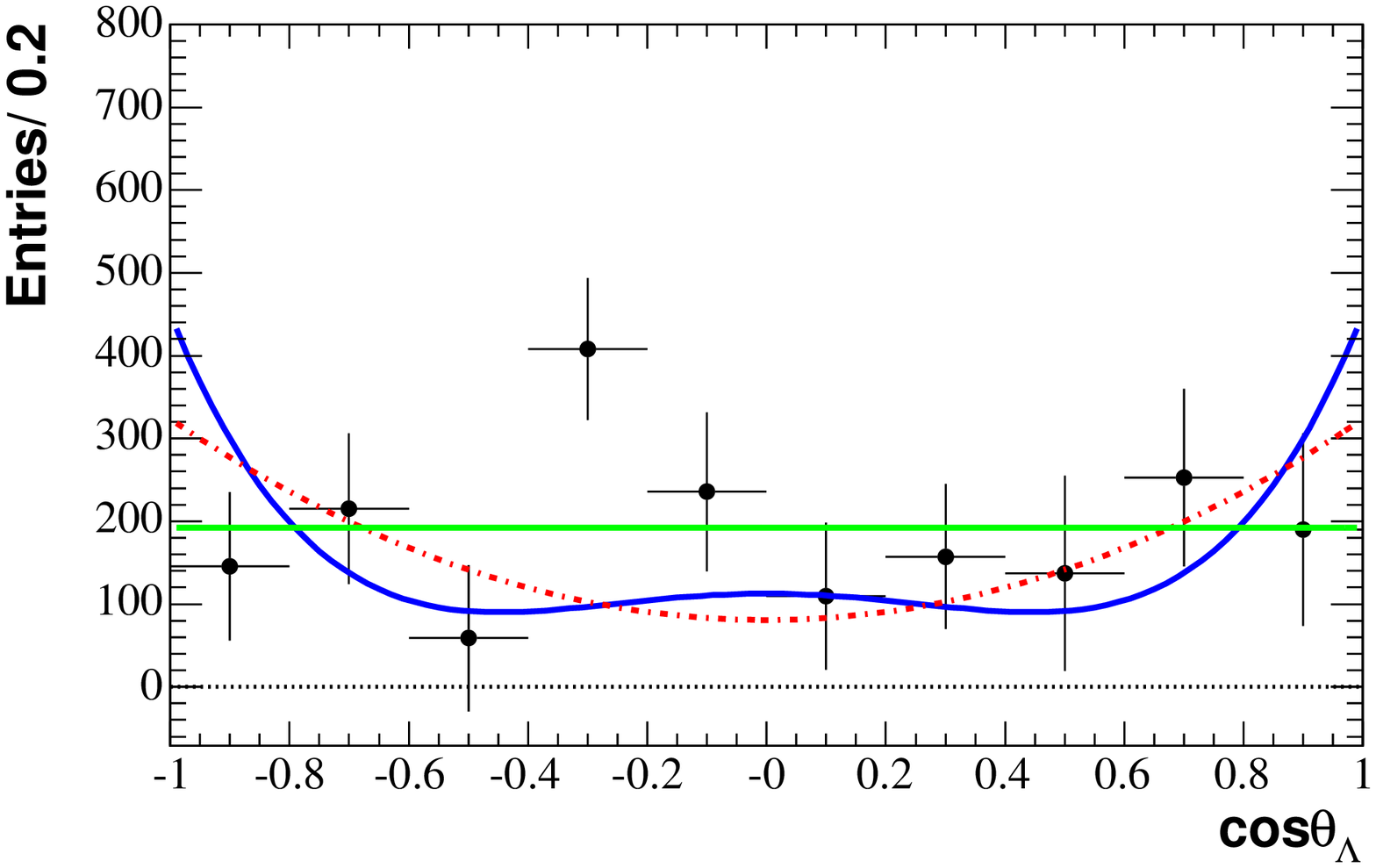}
\begin{picture}(0.,0.)
    \put(-75,312){\normalsize\babar}
    \put(-75,302){Preliminary}
    \put(-75,140){\normalsize\babar}
    \put(-75,130){Preliminary}
    \put(-215,310){\bf{(a)}}
    \put(-215,140){\bf{(b)}}
    \end{picture}
  \caption{The $\Lambda_c^+$ mass-sideband-subtracted, $\Xi(1690)^0$ mass-sideband-subtracted, 
efficiency-corrected cos$\theta_{\Lambda}$ distribution in data.
The horizontal line represents the expected distribution for $J_{\Xi(1690)}=1/2$, 
the dashed curve corresponds to $J_{\Xi(1690)}=3/2$, 
while the solid curve corresponds to $J_{\Xi(1690)}=5/2$. In Fig.~(a) no correction has been made for Dalitz plot
interference effects.  Fig~(b) shows the resulting distribution after correction for Dalitz plot interference effects.}
  \label{fig:EffCorrBinbyBinXi1690SBS}
\end{figure}

\begin{table}[!t]
\begin{center}
\begin{tabular}{c c c c l c l }
\hline \hline
  $J_{\Xi(1690)}$   & & Fit $\chi^2/$NDF & & Fit C.L. & & Comment\\ \hline
   1/2    & &  11.0/9  & &    $0.28$ & & Fig.~6(a), solid line  \\
   3/2    & &  35.7/9    & &   $4\times 10^{-5}$  & & Fig.~6(a),  dashed curve\\
   5/2    & &  42.9/9  & &   $2\times 10^{-6}$  & & Fig.~6(a), solid curve\\  \hline \hline
\end{tabular}
\end{center}
\caption{The ${\rm cos}\theta_{\Lambda}$ angular distribution fit C.L. values of Fig.~6(a) 
corresponding to $\Xi(1690)^0$ spin hypotheses 1/2, 3/2 and 5/2
for  $\Xi(1690)^{0}\rightarrow \Lambda \bar K^0$ data assuming $J_{\Lambda_c}=1/2$.  No correction has been made for Dalitz plot 
interference effects.}
\label{tab:xifits}
\end{table}

\begin{table}[!t]
\begin{center}
\begin{tabular}{c c c c c c l c  }
\hline \hline
  $J_{\Xi(1690)}$   & & Fit $\chi^2/$NDF & & Fit C.L. & & Comment\\ \hline
   1/2    & &  10.8/9  & &    $0.30$ & & Fig.~6(b), solid line  \\
   3/2    & &  19.5/9    & &   $0.02$  & & Fig.~6(b),  dashed curve\\
   5/2    & &  21.7/9  & &   $0.01$  & & Fig.~6(b), solid curve\\  \hline \hline
\end{tabular}
\end{center}
\caption{The ${\rm cos}\theta_{\Lambda}$ angular distribution fit C.L. values of Fig.~6(b)
 corresponding to $\Xi(1690)^0$ spin hypotheses 1/2, 3/2 and 5/2
for  $\Xi(1690)^{0}\rightarrow \Lambda \bar K^0$ data assuming $J_{\Lambda_c}=1/2$, after correction for Dalitz plot
interference effects.}
\label{tab:xifitscor}
\end{table}

\begin{figure}[!t]
  \centering\small
  \includegraphics[height=8cm]{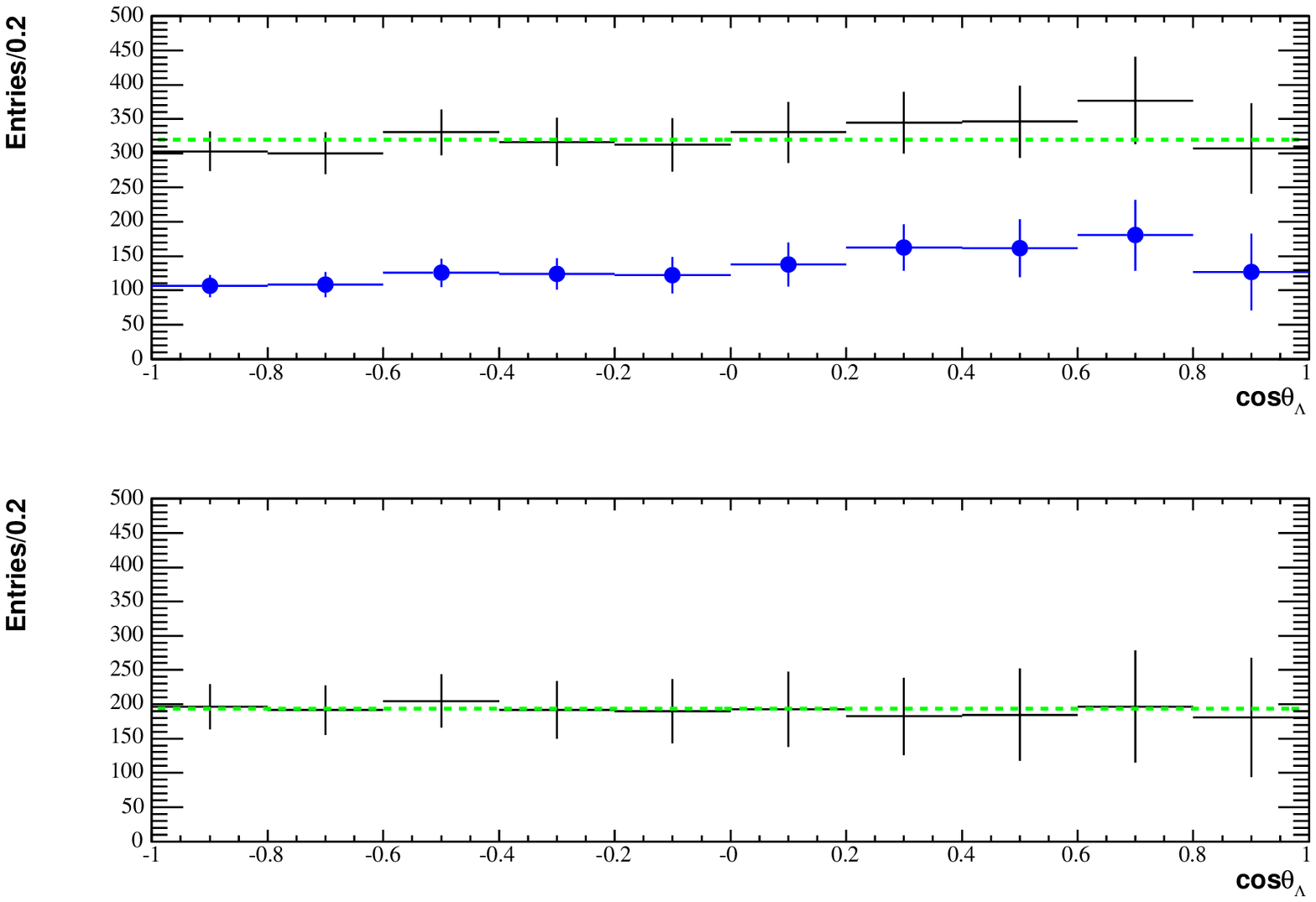}
  \begin{picture}(0.,0.)
 \put(-95,80){\normalsize\babar}
 \put(-55,80){ MC}
    \put(-283,200){\bf{(a)}}
    \put(-283,85){\bf{(b)}}
   \end{picture}
  \caption{The MC 
efficiency-corrected $\Xi(1690)^0$ mass-sideband-subtracted cos$\theta_{\Lambda}$ distribution 
obtained from the $a_0(980)^+$-$\Xi(1690)^0$ Dalitz plot model.  
In Fig.~7(a) the vertical crosses result from the same sideband subtraction procedure applied to data, while 
the contribution from the $\Xi(1690)^0$ background is represented by the solid points.  
In Fig.~7(b) the cos$\theta_{\Lambda}$ distribution for the $\Xi(1690)^0$ only is 
obtained after subtraction of the distributions shown in Fig.~7(a).  
The rise at high cos$\theta_{\Lambda}$ seen in
Fig.~7(a) is due to interference effects in the Dalitz plot.  
The subtraction procedure makes the distribution shown in Fig.~7(b) flat, as expected.}
  \label{fig:EffCorrBinbyBinXi1690SBSimu}
\end{figure}

In Fig.~\ref{fig:EffCorrBinbyBinXi1690SBSimu} the results of treating the MC model of 
the Dalitz plot intensity distribution in 
the same way are shown (where the parameters of the MC model are the ones of Fig.~4). 
As the Monte Carlo sample used contains no spin angular structure, 
it is equivalent to the generation of events with the $(\Lambda \bar{K}^0)$ system in a spin 1/2 (S-wave) state.  
In Fig.~7(a) the upper points correspond to the simulated data 
points of Fig.~6(a), while the solid points result from the contribution to the Dalitz plot 
intensity from the $a_0(980)^+$ and its interference with the $\Xi(1690)^0$.  
Both distributions rise slightly towards cos$\theta_{\Lambda}=1$.  
However, the difference between them, shown in Fig.~7(b), is quite flat as indicated by the dashed line.  The distribution in 
Fig.~7(a) represented by the solid points  
is used to correct the data points of Fig.~6(a) to give the distribution of Fig.~6(b),  
and the new fit results for the different assumptions about the $\Xi(1690)^0$ spin are 
summarized in Table~3.  The spin 1/2 hypothesis is favored, but not as strongly as would be inferred from Table~2.  
This reduced discrimination is due primarily to the fact that the
 background distribution due to $a_0(980)^+$ and its interference with the $\Xi(1690)^0$ (represented by the solid points in Fig.~7(a))
accounts for $\sim 30\%$ of the apparent signal in the $\Xi(1690)^0$ region, 
as shown in Fig.~4, and it is this reduction 
in $\Xi(1690)^0$ signal size which affects the C.L. values for the spin 3/2 and 5/2 hypotheses.

\vfill

\section{CONCLUSIONS}

An analysis of the amplitude structure in the $\Lambda_c^+ \rightarrow \Lambda K_S K^+$ Dalitz plot 
has resulted in the following values for the mass and width of the $\Xi(1690)^0$: 

$$m(\Xi(1690))= 1684.7{\pm 1.3}\;(\rm{stat.})\,^{+2.2}_{-1.6}\;(\rm{syst.})\;\; \rm{MeV}/c^2, $$
$$\Gamma(\Xi(1690))=8.1_{-3.5}^{+3.9}\;(\rm{stat.})\,^{+1.0}_{-0.9}\;(\rm{syst.})\;\; \rm{MeV}.$$
 
The spin of the $\Xi(1690)$ is consistent with 1/2, from fits to the angular 
distribution in data and MC.

\bigskip
\bigskip
\bigskip
\vfill

We are grateful for the 
extraordinary contributions of our \pep2\ colleagues in
achieving the excellent luminosity and machine conditions
that have made this work possible.
The success of this project also relies critically on the 
expertise and dedication of the computing organizations that 
support \babar.
The collaborating institutions wish to thank 
SLAC for its support and the kind hospitality extended to them. 
This work is supported by the
US Department of Energy
and National Science Foundation, the
Natural Sciences and Engineering Research Council (Canada),
Institute of High Energy Physics (China), the
Commissariat \`a l'Energie Atomique and
Institut National de Physique Nucl\'eaire et de Physique des Particules
(France), the
Bundesministerium f\"ur Bildung und Forschung and
Deutsche Forschungsgemeinschaft
(Germany), the
Istituto Nazionale di Fisica Nucleare (Italy),
the Foundation for Fundamental Research on Matter (The Netherlands),
the Research Council of Norway, the
Ministry of Science and Technology of the Russian Federation, and the
Particle Physics and Astronomy Research Council (United Kingdom). 
Individuals have received support from 
the Marie-Curie IEF program (European Union) and
the A. P. Sloan Foundation.


\begin{thebibliography}{99}
                                                                                                                                                  
\bibitem{ref:pdg}
{Particle Data Group,
S. Eidelman {\em et al.}, Phys.\ Lett.\ {\bf B592}, 1 (2004). }
                                                                                                                                                  
\bibitem{ref:r2}
{C. Dionisi {\em et al.}, Phys.\ Lett.\ {\bf B80}, 145 (1978). }
                                                                                                                                                  
\bibitem{ref:r3}
{S.F. Biagi {\em et al.}, Z.\ Phys.\ C {\bf 9}, 305 (1981). }
                                                                                                                                                  
\bibitem{ref:r4}
{S.F. Biagi {\em et al.}, Z.\ Phys.\ C {\bf 34}, 15 (1987). }
                                                                                                                                                  
\bibitem{ref:r5}
{M.I. Adamovich {\em et al.}, Eur.\ Phys.\ J.\ {\bf C5}, 621 (1998). }
                                                                                                                                                  
\bibitem{ref:Bel}
{K. Abe {\em et al.}, Phys.\ Lett.\ {\bf B524}, 33 (2002). }

\bibitem{ref:cc}
{Charge conjugation is implied throughout this paper. }

\bibitem{ref:babar}
 B.\ Aubert {\em et al.},
Nucl.\ Instrum.\ Methods A {\bf 479}, 1 (2002).

\bibitem{ref:Flatte}
{S. M. Flatt\'{e}, Phys.\ Lett.\ {\bf B63}, 224 (1976). }

\bibitem{ref:crystalbar}
{A. Abele {\em et al.}, Phys.\ Rev.\ {\bf D57}, 3860 (1998). }

\bibitem{ref:antimo}
{B. Aubert {\em et al.}, Phys.\ Rev.\ {\bf D72}, 052008 (2005). }

\bibitem{ref:brian}
{B. Aubert {\em et al.}, Phys.\ Rev.\ {\bf D72}, 052006 (2005). }
                                                                                                             
\bibitem{ref:form}
M. Jacob and G. C. Wick, Ann.\ Phys.\ {\bf 7}, 404 (1959).
  
\bibitem{ref:form2}
S. U. Chung, CERN Yellow Report, CERN 71-8 (1971).
  
\bibitem{ref:Eckart}
E. Wigner, Z.\ Phys.\ {\bf 43}, 624 (1927); \\
C. Eckart, Rev.\ Mod.\ Phys.\ {\bf 2}, 305 (1930).
  
\bibitem{ref:fermi}
H. L. Anderson, E. Fermi, E. A. Land and D. E. Nagle, Phys.\ Rev.\ {\bf 85}, 936 (1952).
  
\bibitem{ref:omesp}
B.\ Aubert {\em et al.}, hep-ex/0606039.
 
  
\end{thebibliography}
\end{document}